\documentclass[%
 aps,
 prl,%
 amsmath,amssymb,
 reprint,%
]{revtex4-1}

\usepackage{graphicx}
\usepackage{dcolumn}
\usepackage{bm}
\usepackage{amsmath}

\def \n{{\bf n}}

\def\r{{ \bf r}}
\def\bnabla{{ \bf {\nabla}}}
\begin{document}

\preprint{AIP/123-QED}

\title{Rayleigh breakup of a charged viscous drop via tip-streaming}


\author{Neha Gawande}
\author{Y.S. Mayya}%
\author{Rochish Thaokar}%
\affiliation{ Department of Chemical Engineering, Indian  Institute of  Technology Bombay, Mumbai-400076, India.}

\date{\today}

\begin{abstract}
The experimental observation of D. Duft, T. Achtzehn, R. Muller, B. A. Huber, and T. Leisner, Nature 421, 128 (2003) on the sequential progression of the instability of a charged liquid drop points at the formation of a jet followed by the emission of progeny droplets as a crucial pathway of the Rayleigh fission process.  In spite of considerable theoretical progress, a quantitative understanding of this breakup pathway through  mathematical models has largely remained inconclusive. This limitation has mainly been due to the fact that the generally applied electrostatic boundary condition of the equipotential surface may not be valid near conical ends that experience a singularly fast dynamics near the point of fission. Considering this, we address the problem by invoking the surface charge dynamics within the framework of an axisymmetric boundary element method (BEM), in the viscous limit. The abandonment of the equipotential assumption gives rise to weak tangential electric stresses which turn out to be key the contributors to the emergence of a jet followed by formation of a progeny droplet. The simulations further predict that the size of the progeny droplet follows an inverse power-law scaling relationship with the conductivity of the liquid drop and the smaller sized progenies carry a charge close to its Rayleigh limit.
\end{abstract}

\pacs{Valid PACS appear here}
\maketitle

%
A charged drop of radius $a$ suspended in a medium with electrical permittivity, $\epsilon_e$ undergoes an instability when the total charge on the drop exceeds a critical value of $Q_c=8 \pi \sqrt{\gamma a^3 \epsilon_e}$, where $\gamma$ is the surface tension of the drop \cite{rayleigh1882}. This is termed as Rayleigh instability, which is believed to be responsible for the breakup of raindrops in thunderstorms, the formation of sub-nanometer droplets in electrosprays and generation of ions in ion-mass spectrometry \cite{rosell1994,fenn1989}. This instability occurs when the repulsive Coulombic force overcomes the restoring surface tension force. An infinitesimal quadrupolar shape perturbation (the $2^{nd}$ Legendre mode) on a spherical drop charged beyond $Q_c$ is known to be the most unstable mode \cite{thaokar10}. Although the Rayleigh limit predicts the point of onset of instability, it leaves the details of the break up pathway completely unspecified. The inherent complexity present in the breakup mechanisms was demonstrated only recently \cite{duft03,giglio08} through systematic experiments on a levitated charged drop in a quadrupolar trap. These experiments indicated that above its Rayleigh limit, a charged drop gradually deforms into the shape of a prolate spheroid, elongating further to form sharp conical tips, wherefrom a jet emerges out within a very short time. These jets further disintegrate into a cloud of smaller daughter droplets which eventually take away a significant fraction (20-50\%) of the original charge, although the associated mass loss is small (0.1-1\%) \cite {doyle64, abbas67, roulleau72, richardson89, taflin89, duft03}. The sizes and charge on the daughter droplets thus formed are important since they determine whether the progenies can undergo further breakup or not. In this letter, we focus on providing a model for explaining the observed break-up pathway and estimation of the size and charge on the daughter droplets.  

 The Rayleigh fission process of an isolated charged drop is generally modeled under the assumption of a perfectly conducting (PC) liquid drop in which the charges are distributed uniformly on its equipotential surface. The flow equations are solved numerically using boundary element method (BEM) either in the viscous flow limit \cite{betelu06} or in the potential flow limit \cite{burton11}. Both these studies show that the charged drop deforms initially into the shape of a spheroid, progressively deforming into an elongated object with sharp conical tips, whereafter it undergoes a numerical singularity. The results in the viscous flow limit indicate that the capillary stresses at the sharp tips become subdominant and a balance of the viscous and the electric stresses leads to the formation of a dynamic cone angle of about $25^{o}$. In contrast, the simulations for the potential flow limit yield a cone angle of about $49.3^{o}$, coincidentally close to the classical equilibrium angle of the Taylor cone derived from static considerations. Actual experimental images show a cone angle of $30^o$ indicating significant viscous effects \cite{giglio08}. Furthermore, the  PC model was also used for predicting fractional charge loss of about 39\% \cite{gawande2017} assuming negligible mass loss. To proceed beyond singularity within the framework of PC model and predict ejection, \citet{garzon14} performed  BEM simulations coupled with a level set technique for  inviscid drops.  Although the model could predict daughter droplets, the ejection occurred from protrusions, whose lengths were far smaller (1/5th of the droplet diameter), in contrast to the experimental observation of long (3 times the drop diameter) jets \cite{duft03, giglio08}. Besides, in view of the absence of viscosity, these protrusions were likely to be artefacts of inertial excursions rather than due to sustained tangential stresses necessarily required for jet formation. Considering these, the PC model fails to predict jet formation and one needs to look for alternative mechanisms to explain the complex pathway of break up seen in experiments. \citet{collins08, collins13} had observed that charge dynamics and viscous stresses are necessary for the jet formation in the study of breakup of uncharged oil drops having low conductivity {\em{under strong applied electric fields}}. Taking a clue from this, we apply it to the case of Rayleigh break up of a charged drop ({\em{in the absence of any external electric field}}) with high but finite conductivity which involves faster dynamics and high electrical stresses in the viscous limit.

In this spirit the problem is solved numerically by considering an electrically charged drop of a conducting liquid of viscosity ${\mu}_i$, density ${\rho}_i$ and the conductivity $\sigma_i$ suspended in a perfectly dielectric Newtonian fluid medium of viscosity $\mu_e$ with a permittivity, ${\epsilon}_e$. The external medium is considered to be a gas or air. The electric potential on the surface of the drop due to presence of surface charge is denoted by $(\tilde {\phi})$ and the electric field is expressed as $\tilde{{\bf E}}=-\tilde{{\bf \nabla}} (\tilde {\phi})$, with $\tilde{\nabla}^2 \tilde{\phi}=0$. We consider the hydrodynamics in the Stokes flow limit, such that the Ohnesorge number, $Oh=\mu_i/\sqrt{\rho_i a \gamma}$, is large. The dimensional quantities are indicated by tilde and non dimensional quantities are without tilde. The nondimensional parameters used in this problem are as follows: length scales are of the order of the initial radius $a$ of the spherical drop. The time is nondimensionalized by the hydrodynamic timescale, $t_h=\mu_i a/\gamma$, the velocity by $\gamma/\mu_i$, the surface charge density $\tilde q$ by $\sqrt{\frac{\gamma \epsilon_e}{a}}$ and the electric field by $\sqrt{\frac{\gamma}{a \epsilon_e}}$. The total surface charge is nondimensionalized by $\sqrt{\gamma a^3 \epsilon_e}$ such that the non dimensional Rayleigh charge is $8\pi$.  The electrostatics and the Stokes equations are solved using the axisymmetric boundary integral method, using well established methodologies \cite{deshmukh12,lanauze2015,gawande2017}.

 For a finitely conducting (FC) charged drop the electric boundary conditions at the interface in the scaled variables can be written as, $E_{ne}-S E_{ni}=q$, ${E_{t_e}}={ E_{t_i}}$. Thus the boundary integral equation for the electric field calculation is given by,
 
 \begin{multline}
  \frac{(S+1)}{2S} E_{n_e}(\r_s)+\frac{1}{4\pi}(\frac{S-1}{S})\int {\bf{n}}\cdot \bnabla {\bf{G}}^e(\r,\r_s) E_{n_e}(\r) dA(\r) \\ 
  =\frac{1}{2S}q(\r)-\frac{1}{4\pi S}\int {\bf{n}}\cdot\bnabla {\bf{G}}^e(\r,\r_s) q(\r)dA(\r)
  \label{eqn:BIforE}
 \end{multline}
 and for the electrostatic potential $\phi(\r_s)$,
 \begin{equation}
 \phi(\r_s)=\frac{1}{4\pi} \int {\bf{G}}^e(\r,\r_s) (E_{n_e}(\r)-E_{n_i}(\r))dA(\r)
 \label{eqn:BIforpot}
 \end{equation}
 where ${\bf{G}}^e (\r,\r_s)=\frac{1}{ |{\bf r-r_s}|}$ while ${\bf{r}}$ and ${\bf {r}_s}$ are the position vectors on the surface of the drop with area $A$ and $E_{ne}=\bf{E_e}\cdot \n$ where $\n$ is the outward unit normal. The conservation of the total surface charge is ensured through the charge dynamics equation which on non-dimensionalisation reduces to,
\begin{equation}
\frac{\partial q}{\partial t}=\frac{S}{Sa}{\bf{E_{ni}}}- \left(\frac{1}{r}\Bigr{[}\frac{\partial}{\partial s}(q r v_t)\Bigr{]}+q(\bnabla_s \cdot \bf{n})(\bf{v}\cdot \bf{n}) \right)
\label{eqn:charge_conservation}
\end{equation}

where, $S=\epsilon_i/\epsilon_e$ is the ratio between permittivities of the drop and the external medium while $Sa=t_e/t_h$, is the nondimensional number known as Saville number where, $t_e=\epsilon_i/\sigma_i$ is the charge relaxation time and $t_h$ is the hydrodynamic timescale. $\bnabla_s=(\bf{I}-\bf{n}\bf{n})\cdot \bnabla$ represents the surface gradient \cite{stone90}. The first term on the right hand side of the equation \ref{eqn:charge_conservation} accounts for charges brought to the surface by conduction, while second and third terms are convection terms. The second term represents the meridional advection of charges while the third term is a source like term which accounts for local change in the charge density due to dilation of the drop surface. Here, the outside electric field does not appear as the conductivity of the external fluid medium, $\sigma_e$, is considered to be zero. The force density responsible for drop deformation is then given by, $\triangle{\bf{f}}= {\bf{n}}[\bnabla \cdot {\bf{n}}-[{\bf{\tau}^e}]] $, where $[{\bf{\tau}^e}]$ is the nondimensional jump in the electrical traction across the interface and is given by,
\begin{equation}
[{\bf{\tau}^e}]=\frac{1}{2}[(E_{n_e}^2-SE_{n_i}^2)+(S-1)E_{t_e}^2]\n+q E_{t_e} \bf{t}
\label{eqn:taue}
\end{equation}
  To initiate the evolution process the drop is deformed slightly into a shape of a prolate spheroid and a total charge of $8.1\pi$ (which is slightly above the Rayleigh limit) is then distributed uniformly on the deformed drop. In the present simulations it is ensured that the total surface charge is conserved to an accuracy of 1\%. Numerically an adaptive meshing is used to ensure that the local grid size $\triangle s_{min}$ is always smaller than the minimum neck radius. The time steps are adapted using the criteria, $\triangle t= C \triangle s_{min}/v_{n_{max}}$, where $C$ denotes the CFL number which is kept constant at 0.01 and $v_{n_{max}}$ is maximum velocity with which the grid points move in the given timestep (see SM for details).
   
  \begin{figure}
 \centering
     \includegraphics[width=0.45\textwidth]{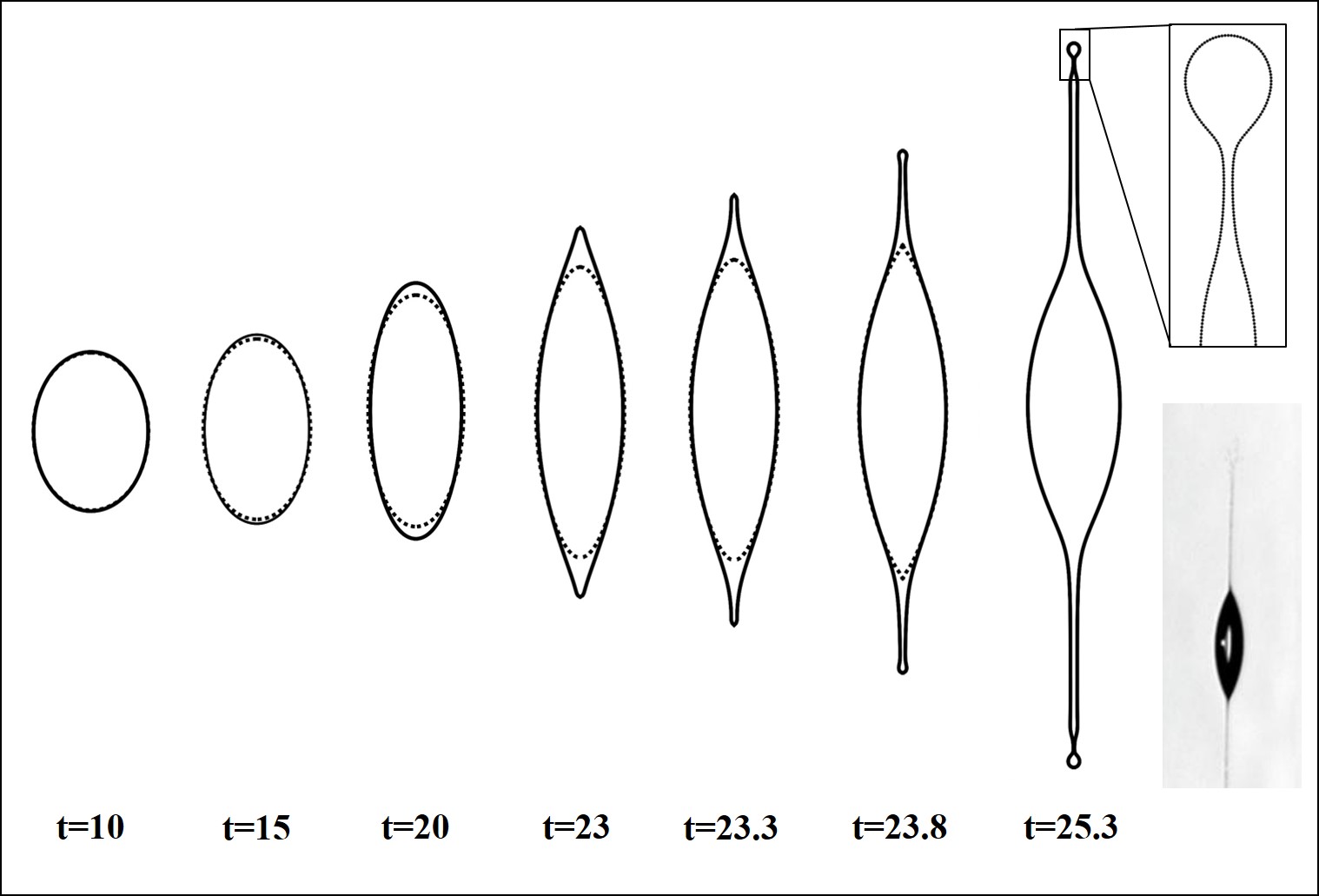}  
   \caption{Comparison of the temporal evolution of the drop shapes for the two cases of PC (dotted line) and FC (solid line) drop models. PC drop model forms sharp conical ends at t=23.8 and exhibits numerical singularity, while FC drop model continues to deform further and jet is ejected out with a small progeny at the tip of the drop. The inset at the right bottom corner is the experimental image of drop breakup presented by \citet{duft03} (reprinted with permission).}
   \label{fig:sequence}
 \end{figure}
 
 The effect of conductivity is introduced through a $Sa$. Typically, for example, a methanol drop of radius 50 $\mu m$ size, with the conductivity ($\sigma_i=4\times10^{-4} S/m$) taken at room temperature has $Sa=0.55$. This indicates that, when the length scales are of the order of droplet radius $a$, the charge relaxation is faster than the characteristics timescales used in the simulations. Thus it appears that the PC drop model may suffice to predict the Rayleigh fission process. For the PC drop, the charge distribution is instantaneous and the inside electric field is zero due to the assumption of an equipotential surface. Thus for PC drop, equation \ref{eqn:BIforpot} is used to calculate the external electric field $E_{ne}$ by putting $E_{ni}=0$. The unknown potential $\phi({\bf r_s})$ is constant on the surface of the drop, and is determined by the condition of conservation of charge, $\int E_{ne} ({\bf r}) dA({\bf r})=Q$, where $Q$ is the initial charge on the drop surface which is conserved during the shape evolution. Thus the nondimensional jump in the electrical stresses in PC drop model is given by, $[{\bf{\tau}}^e]=\frac{1}{2} E_{ne}^2 \n$.


\begin{figure}
\centering
\includegraphics[width=0.48\textwidth]{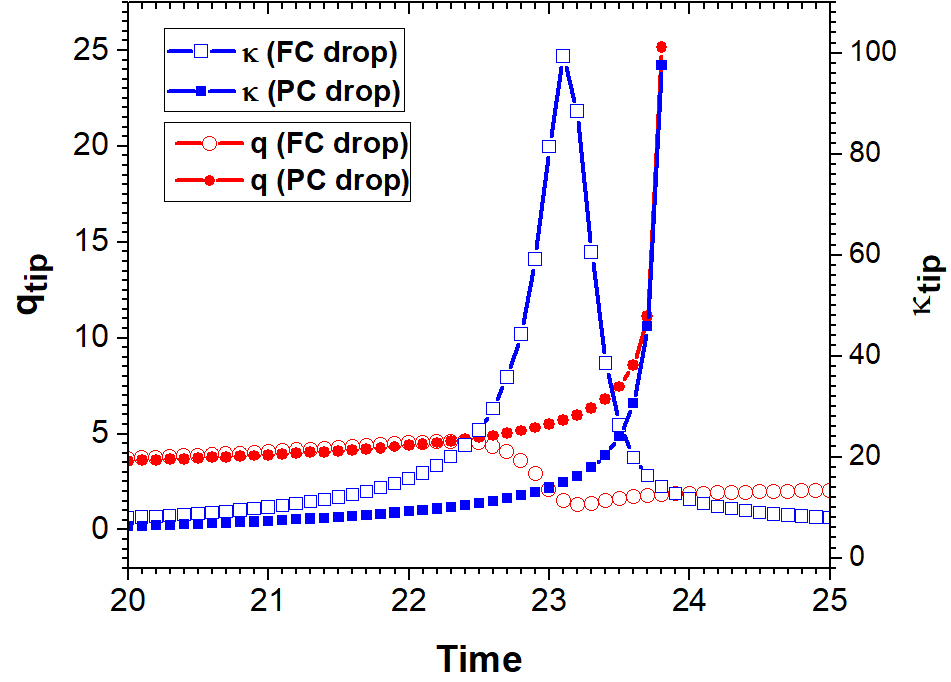}
\caption{Comparison of the temporal evolution of curvature and charge density at the tip of the drop for the two cases of PC (filled symbols) and FC (open symbols) drop models.}
\label{fig:tvskandq}
\end{figure}

\begin{figure}
\includegraphics[width=0.48\textwidth]{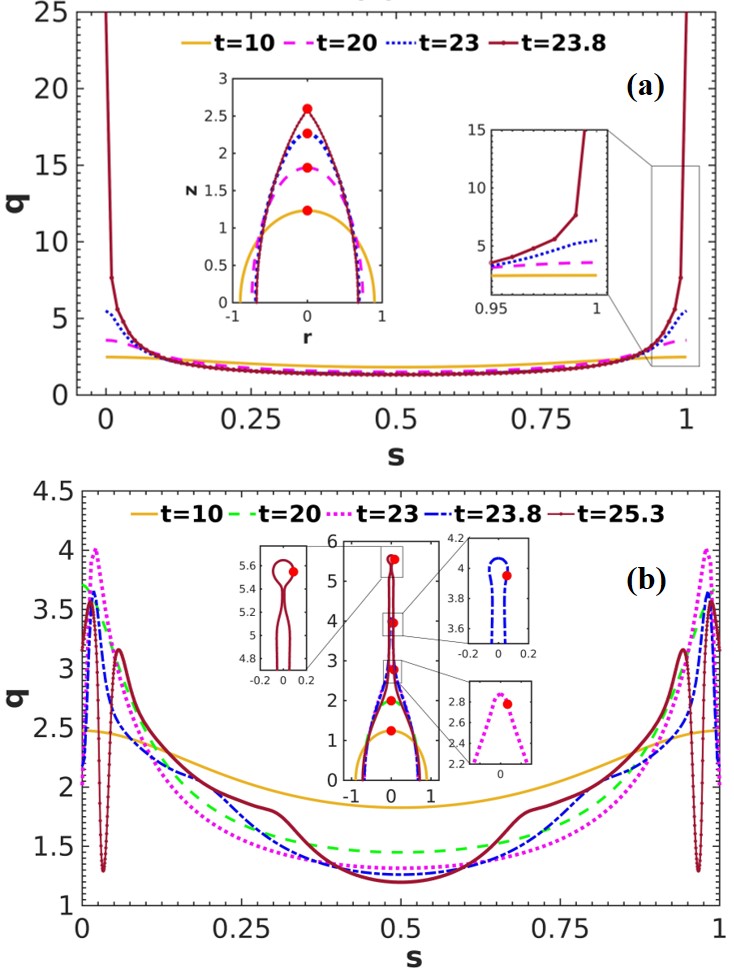} 
\caption{The distribution of charge density on the drop surface as a function of normalised arclength ($s$) at various times and drop shapes indicating point of maximum charge density (red dot) for the corresponding times in case of (a) PC and (b) FC drop model. For FC drop maximum charge density shifts from the poles of the drop towards the equator with time. Insets show the magnified figures near the poles for better clarity.}
\label{fig:chargedensity}
\end{figure}

The typical drop deformation sequences with time in PC and FC drop models are shown in figure \ref{fig:sequence}. At $t=23.8$, the PC drop model exhibits a shape singularity owing to its limitation of instantaneous charge transport and the absence of tangential electric stresses. Precisely this limitation is overcome by the FC model in which the finite time taken by the flow of charge to the regions of high curvature delimits the build up of charge at the tip of the drop to a finite value since by then the capillary stresses relax the tip curvature ($\kappa_{tip}$) and the simulations can be continued further. A temporal analysis shown in figure \ref{fig:tvskandq} indicates that the charge density at the poles seems to reduce earlier in time (at $t$=22.5) than the curvature (maxima at $t$=23.2), suggesting that the reduction in curvature at the poles is a consequence of the reduction in charge density. At this stage the electric potential near the tip of the FC drop reduces and equipotential assumption is no more valid (refer supplementary material for potential distribution). The spatial variation of charge density and the curvature as shown in figure \ref{fig:chargedensity}(b) indicate that these variables are now extremized at a location below the poles unlike the PC drop where the charge density and curvature remains maximum at the tip of the drop (figure \ref{fig:chargedensity}(a)).

\begin{figure}
\centering
    \includegraphics[width=0.48\textwidth]{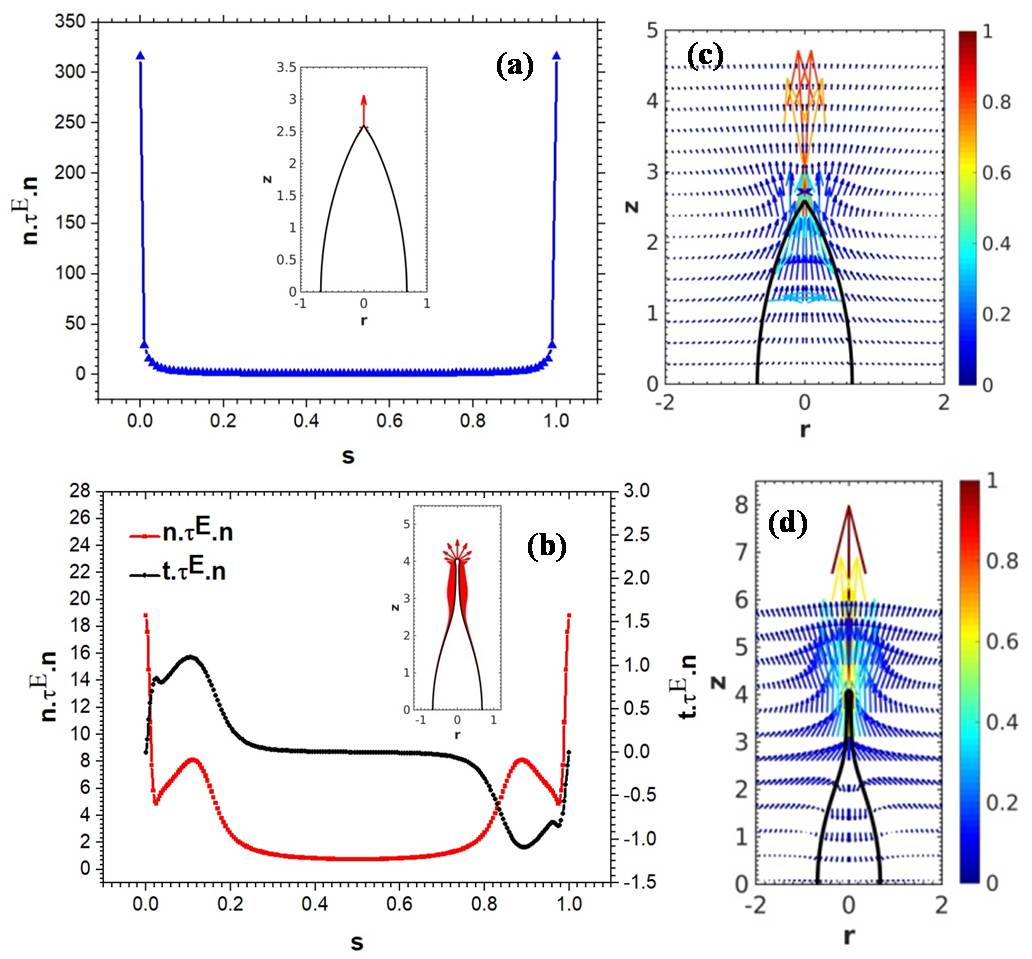}    
  \caption{Electric stress distribution and velocity profiles in case of (a),(b) PC drop and (c), (d) FC drop model at time t=23.8. The electric stress is purely normal in case of PC drop model with maximum stress acting on the tip of the drop while stress distribution is modified due to presence of weak tangential stresses in case of FC drop model. The velocity profiles show the reversal of flow due to modification of stress distribution in FC drop model.}
	\label{fig:pcfc_stress}
\end{figure}

 The reason for deviation from the equipotential state is indeed the finite charge dynamics admitted by the FC model. As the dynamics accelerates after the formation of conical ends, the length scale independent charge dynamics ($t_e$) becomes comparable to the size ($l$) dependent hydrodynamic time scale ($t_{hl}=\mu_i l/\gamma$). While in a slightly deformed drop (upto the formation of conical ends), the length scale can be assumed to be of the order of the size of the drop, subsequently, the curvature at the poles becomes the more relevant length scale. The slow charge dynamics relative to the hydrodynamics now means that the charges cannot reach instantaneously to the new surface created, resulting in violation of the equipotential assumption. The variation of charge density and potential along the surface of the drop leads to tangential field and thereby tangential electric stresses. Unlike the normal electric stresses which can be balanced by the capillary forces, the tangential electric stresses lead to tangential fluid flow in the system. Thus a hyperboloidal tip is formed in the FC drop from where a jet emerges out (figure \ref{fig:sequence}). 

At the time of formation of conical tips the normal electric stresses are maximum at the poles in the PC model (figure \ref{fig:pcfc_stress}(a)). 
On the other hand, in the FC model, the normal electric stress increases with time upto the formation of conical ends, but subsequently shows a dramatic reduction at the poles. The tangential stress on the other hand, while nearly negligible upto the cone formation shows a buildup with time. In the PC model the pressure (see SM for pressure distribution) at the poles is negative due to large normal tensile electric stress, thereby leading to a parabolic axial velocity profile as shown in figure \ref{fig:pcfc_stress}(c). In the FC model, the pressure at the poles and in the neck region can be positive and high. The tangential stresses and the pressure distribution then leads to an axial velocity profile which is maximum at the drop interface in the jet region (figure\ref{fig:pcfc_stress}(d)). This causes a flow reversal inside the droplet. Thus the modification of normal electric stress distribution due to the presence of tangential electric stress leads to emergence and subsequent fattening of a jet from the conical ends of the droplet. The role of tangential stresses is affirmed by switching them off in the force balance and the jet formation is not observed. The charge dynamics has three contributions, conduction from the bulk, convection along the surface and change in charge density due to surface dilation. Our analysis indicates that the charge density and jet formation is most affected by the surface dilation terms. The high stretching of the interface due to normal forces at the poles, leads to depletion of charges thereby kick-starting the formation of a jet and its subsequent necking (see SM for more details).

\begin{figure}
\includegraphics[width=0.49\textwidth]{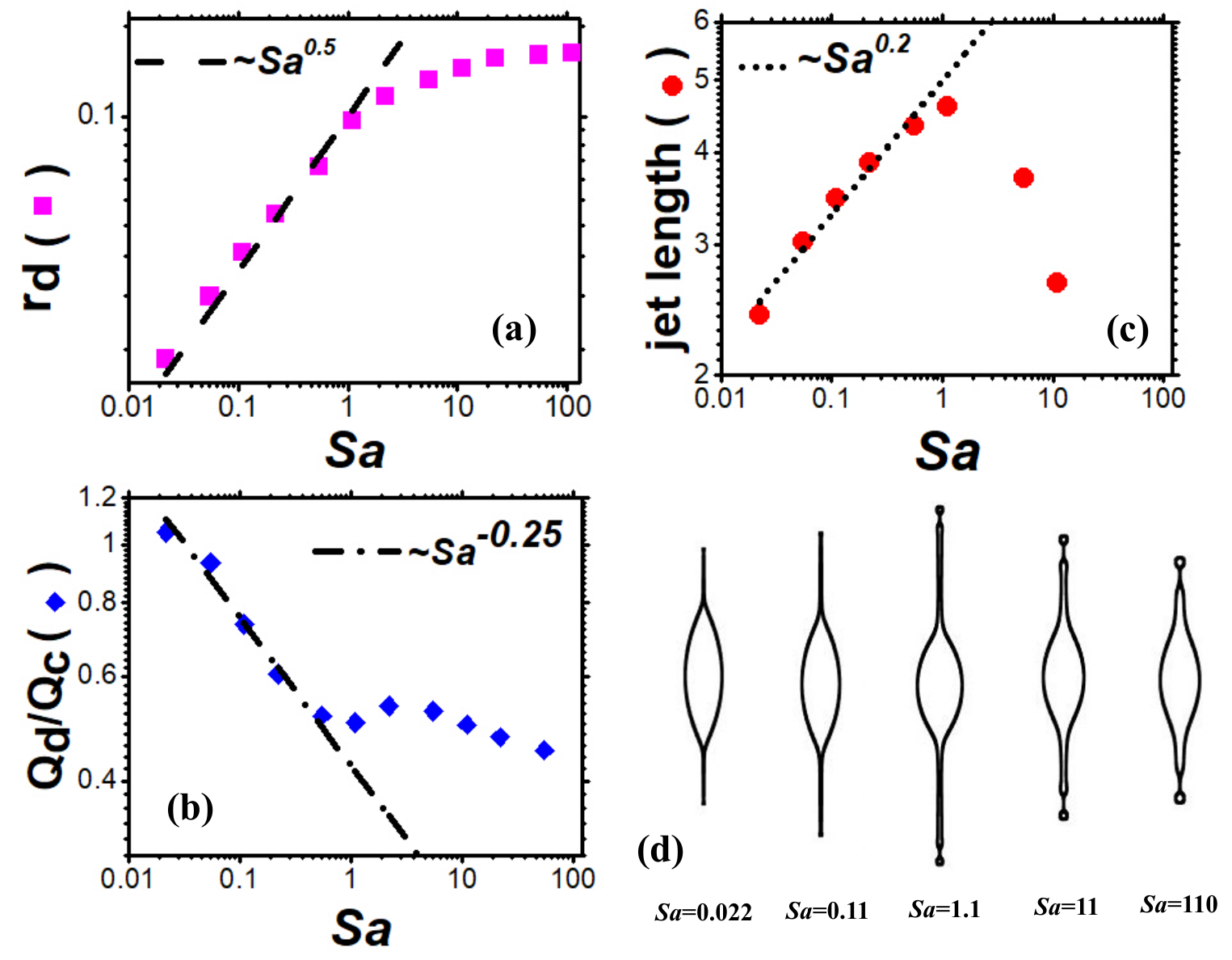}
\caption{Effect of Sa on the drop shapes formed before breakup in FC drop case. (a)Size of the progeny, (b)length of the jet, (c)the ratio of charge carried by the progeny droplet to its Rayleigh limit indicates that the progeny droplets are unstable at the time of their formation and (d) Deformed drop shapes at the onset of pinch off of progeny as a function of $Sa$.}
\label{fig:rdqdscale}
\end{figure}	

Figure \ref{fig:rdqdscale}(a) shows the effect of conductivity in terms of $Sa$ on the size of the progeny formed during the Rayleigh breakup. It is observed that the radius of progeny ($r_d$) formed is lower for lower $Sa$ which implies that a liquid drop of higher conductivity will form smaller progenies. This is in agreement with the previous studies \cite{burton11,collins13}. A naive scaling of a balance of the electric time scale $t_e$ and the hydrodynamic time scale $t_{hl}$ leads to $l/a\sim Sa$. On the other hand if we consider that the jet is issued after the conical tips approach the singularity, we find that radius of the jet ($r_j$) is equivalent to the reciprocal of the curvature ($1/\kappa$) at the tip of the drop that scales as $(t_o-t)^{1/2}$ (refer \cite{gawande2017}). In dimensional terms, this suggests that $ r_j/a\sim [(\tilde{t}_o-\tilde{t})/(\mu_i a/\gamma)]^{1/2}$. Realising that the charge loss occurs over the electric time scale $(\tilde{t}_o-\tilde{t})\sim t_e$, leads to $r_j/a\sim Sa^{1/2}$. Thus over the length scale $l$, the jet has a lower charge and thereby the surface tension forces become dominant in the jet region. This leads to jet breakup by the Rayleigh Plateau instability and forms the progeny droplets of size equivalent to the radius of the jet. This qualitatively explains the progeny droplet size $r_d\sim Sa^{1/2}$ as observed in the simulations. Similarly, the scalings (shown in figure \ref{fig:rdqdscale}(c)) for the dimensional charge present on the progeny can be explained by singular scaling of charge density at the incipience of a jet which is given by, $q_d\sim [(t_o-t)/(\mu_i a/\gamma)]^{-1/2}$ \cite{gawande2017}. Thus the total charge on the daughter droplet, $Q_d\sim q_d {r_d}^2$  implies that $Q_d\sim Sa^{1/2}$  over the electric timescale $t_e$. This when presented in terms of the fraction of the Rayleigh charge ($Q_c$) results in $\tilde{Q}_d\sim Q_c Sa^{-1/4}$ (figure \ref{fig:rdqdscale} (b)). This indicates that the Rayleigh fission of a charged droplet with high conductivity will produce the marginally stable progeny droplets. This result is in agreement with the results predicted by potential flow analysis \cite{burton11}. The asymptotic results of high $Sa$ are in agreement with perfect dielectric calculations (not discussed here) which are independent of the conductivity of the droplet. Thus, the weak scaling of $r_d$ ($\sim Sa^{0.1}$) at higher Sa can be attributed to strong dielectric effects. It is also observed that the jet length increases with the decrease in conductivity and reaches to a maximum value for $Sa=1.1$ but reduces for higher $Sa$ values (figure \ref{fig:rdqdscale}(c)). The drop shapes at the onset of breakup shown in figure \ref{fig:rdqdscale}(d) indicate that the drops with higher conductivity form a distinct jet before a progeny detaches from the tip of the drop by pinch-off. Our numerical analysis indicates that, while the jet incipience occurs over a fast time scale of $Sa$, the jet elongates over a slower time scale of $Sa^{1/2}$. Since, the jet velocity scales as $Sa^{-1/3}$ it leads to jet length scaling as $Sa^{1/6}$. However, at lower conductivities the dominance of capillary stresses occurs much earlier than the formation of sustained jet and the droplet breaks by pinch off. The high $Sa$ is not of much practical relevance in studies on Rayleigh breakup and electrosprays wherein salts are often added to increase the conductivity of the liquids.  \\

We have investigated the formation of daughter droplets due to a highly nonlinear breakup of a charged jet, in the viscous limit. The analysis is valid when $Oh\gg1$. Thus the present analysis could be considered for the breakup of droplets of sizes, of the order of their viscous length scales ($\mu_i^2/(\rho \gamma)$) or smaller. For example, the results presented in this work will hold for the case of $4 \mu m$ droplets for $1-Octanol$, $6 \mu m$ droplet for n-Decanol or $32 \mu m$ droplet for 3-Ethylene glycol. We propose that since droplets in the processes such as electrospray or ionisation in ion mass spectroscopy, eventually undergo Rayleigh fission at the smallest length scales, the viscous analysis does become relevant in these processes at late stages, and could actually explain the nanometer sized daughter droplets formed in some experiments on electrospray \cite{chen1995,singh2016}. Thus while the analysis of \citet{collins13} will indeed hold good for prediction of the droplet size emerging from a Taylor cone in an electrospray experiment under an applied electric field, the final size distribution could be governed by the viscous scaling suggested in this work. 

\bibliography{refss2.bib}

\begin{thebibliography}{22}%
\makeatletter
\providecommand \@ifxundefined [1]{%
 \@ifx{#1\undefined}
}%
\providecommand \@ifnum [1]{%
 \ifnum #1\expandafter \@firstoftwo
 \else \expandafter \@secondoftwo
 \fi
}%
\providecommand \@ifx [1]{%
 \ifx #1\expandafter \@firstoftwo
 \else \expandafter \@secondoftwo
 \fi
}%
\providecommand \natexlab [1]{#1}%
\providecommand \enquote  [1]{``#1''}%
\providecommand \bibnamefont  [1]{#1}%
\providecommand \bibfnamefont [1]{#1}%
\providecommand \citenamefont [1]{#1}%
\providecommand \href@noop [0]{\@secondoftwo}%
\providecommand \href [0]{\begingroup \@sanitize@url \@href}%
\providecommand \@href[1]{\@@startlink{#1}\@@href}%
\providecommand \@@href[1]{\endgroup#1\@@endlink}%
\providecommand \@sanitize@url [0]{\catcode `\\12\catcode `\$12\catcode
  `\&12\catcode `\#12\catcode `\^12\catcode `\_12\catcode `\%12\relax}%
\providecommand \@@startlink[1]{}%
\providecommand \@@endlink[0]{}%
\providecommand \url  [0]{\begingroup\@sanitize@url \@url }%
\providecommand \@url [1]{\endgroup\@href {#1}{\urlprefix }}%
\providecommand \urlprefix  [0]{URL }%
\providecommand \Eprint [0]{\href }%
\providecommand \doibase [0]{http://dx.doi.org/}%
\providecommand \selectlanguage [0]{\@gobble}%
\providecommand \bibinfo  [0]{\@secondoftwo}%
\providecommand \bibfield  [0]{\@secondoftwo}%
\providecommand \translation [1]{[#1]}%
\providecommand \BibitemOpen [0]{}%
\providecommand \bibitemStop [0]{}%
\providecommand \bibitemNoStop [0]{.\EOS\space}%
\providecommand \EOS [0]{\spacefactor3000\relax}%
\providecommand \BibitemShut  [1]{\csname bibitem#1\endcsname}%
\let\auto@bib@innerbib\@empty
\bibitem [{\citenamefont {Rayleigh}(1882)}]{rayleigh1882}%
  \BibitemOpen
  \bibfield  {author} {\bibinfo {author} {\bibfnamefont {L.}~\bibnamefont
  {Rayleigh}},\ }\href@noop {} {\bibfield  {journal} {\bibinfo  {journal} {The
  London, Edinburgh, and Dublin Philosophical Magazine and Journal of Science}\
  }\textbf {\bibinfo {volume} {14}},\ \bibinfo {pages} {184} (\bibinfo {year}
  {1882})}\BibitemShut {NoStop}%
\bibitem [{\citenamefont {Rosell-Llompart}\ and\ \citenamefont
  {De~La~Mora}(1994)}]{rosell1994}%
  \BibitemOpen
  \bibfield  {author} {\bibinfo {author} {\bibfnamefont {J.}~\bibnamefont
  {Rosell-Llompart}}\ and\ \bibinfo {author} {\bibfnamefont {J.~F.}\
  \bibnamefont {De~La~Mora}},\ }\href@noop {} {\bibfield  {journal} {\bibinfo
  {journal} {Journal of Aerosol Science}\ }\textbf {\bibinfo {volume} {25}},\
  \bibinfo {pages} {1093} (\bibinfo {year} {1994})}\BibitemShut {NoStop}%
\bibitem [{\citenamefont {Fenn}\ \emph {et~al.}(1989)\citenamefont {Fenn},
  \citenamefont {Mann}, \citenamefont {Meng}, \citenamefont {Wong},\ and\
  \citenamefont {Whitehouse}}]{fenn1989}%
  \BibitemOpen
  \bibfield  {author} {\bibinfo {author} {\bibfnamefont {J.~B.}\ \bibnamefont
  {Fenn}}, \bibinfo {author} {\bibfnamefont {M.}~\bibnamefont {Mann}}, \bibinfo
  {author} {\bibfnamefont {C.~K.}\ \bibnamefont {Meng}}, \bibinfo {author}
  {\bibfnamefont {S.~F.}\ \bibnamefont {Wong}}, \ and\ \bibinfo {author}
  {\bibfnamefont {C.~M.}\ \bibnamefont {Whitehouse}},\ }\href@noop {}
  {\bibfield  {journal} {\bibinfo  {journal} {Science}\ }\textbf {\bibinfo
  {volume} {246}},\ \bibinfo {pages} {64} (\bibinfo {year} {1989})}\BibitemShut
  {NoStop}%
\bibitem [{\citenamefont {Thaokar}\ and\ \citenamefont
  {Deshmukh}(2010)}]{thaokar10}%
  \BibitemOpen
  \bibfield  {author} {\bibinfo {author} {\bibfnamefont {R.}~\bibnamefont
  {Thaokar}}\ and\ \bibinfo {author} {\bibfnamefont {S.}~\bibnamefont
  {Deshmukh}},\ }\href@noop {} {\bibfield  {journal} {\bibinfo  {journal}
  {Physics of Fluids}\ }\textbf {\bibinfo {volume} {22}},\ \bibinfo {pages}
  {034107} (\bibinfo {year} {2010})}\BibitemShut {NoStop}%
\bibitem [{\citenamefont {Duft}\ \emph {et~al.}(2003)\citenamefont {Duft},
  \citenamefont {Achtzehn}, \citenamefont {M{\"u}ller}, \citenamefont {Huber},\
  and\ \citenamefont {Leisner}}]{duft03}%
  \BibitemOpen
  \bibfield  {author} {\bibinfo {author} {\bibfnamefont {D.}~\bibnamefont
  {Duft}}, \bibinfo {author} {\bibfnamefont {T.}~\bibnamefont {Achtzehn}},
  \bibinfo {author} {\bibfnamefont {R.}~\bibnamefont {M{\"u}ller}}, \bibinfo
  {author} {\bibfnamefont {B.~A.}\ \bibnamefont {Huber}}, \ and\ \bibinfo
  {author} {\bibfnamefont {T.}~\bibnamefont {Leisner}},\ }\href@noop {}
  {\bibfield  {journal} {\bibinfo  {journal} {Nature}\ }\textbf {\bibinfo
  {volume} {421}},\ \bibinfo {pages} {128} (\bibinfo {year}
  {2003})}\BibitemShut {NoStop}%
\bibitem [{\citenamefont {Giglio}\ \emph {et~al.}(2008)\citenamefont {Giglio},
  \citenamefont {Gervais}, \citenamefont {Rangama}, \citenamefont {Manil},
  \citenamefont {Huber}, \citenamefont {Duft}, \citenamefont {M\"uller},
  \citenamefont {Leisner},\ and\ \citenamefont {Guet}}]{giglio08}%
  \BibitemOpen
  \bibfield  {author} {\bibinfo {author} {\bibfnamefont {E.}~\bibnamefont
  {Giglio}}, \bibinfo {author} {\bibfnamefont {B.}~\bibnamefont {Gervais}},
  \bibinfo {author} {\bibfnamefont {J.}~\bibnamefont {Rangama}}, \bibinfo
  {author} {\bibfnamefont {B.}~\bibnamefont {Manil}}, \bibinfo {author}
  {\bibfnamefont {B.~A.}\ \bibnamefont {Huber}}, \bibinfo {author}
  {\bibfnamefont {D.}~\bibnamefont {Duft}}, \bibinfo {author} {\bibfnamefont
  {R.}~\bibnamefont {M\"uller}}, \bibinfo {author} {\bibfnamefont
  {T.}~\bibnamefont {Leisner}}, \ and\ \bibinfo {author} {\bibfnamefont
  {C.}~\bibnamefont {Guet}},\ }\href@noop {} {\bibfield  {journal} {\bibinfo
  {journal} {Physical Review E}\ }\textbf {\bibinfo {volume} {77}},\ \bibinfo
  {pages} {036319} (\bibinfo {year} {2008})}\BibitemShut {NoStop}%
\bibitem [{\citenamefont {Doyle}\ \emph {et~al.}(1964)\citenamefont {Doyle},
  \citenamefont {Moffett},\ and\ \citenamefont {Vonnegut}}]{doyle64}%
  \BibitemOpen
  \bibfield  {author} {\bibinfo {author} {\bibfnamefont {A.}~\bibnamefont
  {Doyle}}, \bibinfo {author} {\bibfnamefont {D.~R.}\ \bibnamefont {Moffett}},
  \ and\ \bibinfo {author} {\bibfnamefont {B.}~\bibnamefont {Vonnegut}},\
  }\href@noop {} {\bibfield  {journal} {\bibinfo  {journal} {Journal of Colloid
  Science}\ }\textbf {\bibinfo {volume} {19}},\ \bibinfo {pages} {136}
  (\bibinfo {year} {1964})}\BibitemShut {NoStop}%
\bibitem [{\citenamefont {Abbas}\ and\ \citenamefont {Latham}(1967)}]{abbas67}%
  \BibitemOpen
  \bibfield  {author} {\bibinfo {author} {\bibfnamefont {M.}~\bibnamefont
  {Abbas}}\ and\ \bibinfo {author} {\bibfnamefont {J.}~\bibnamefont {Latham}},\
  }\href@noop {} {\bibfield  {journal} {\bibinfo  {journal} {Journal of Fluid
  Mechanics}\ }\textbf {\bibinfo {volume} {30}},\ \bibinfo {pages} {663}
  (\bibinfo {year} {1967})}\BibitemShut {NoStop}%
\bibitem [{\citenamefont {Roulleau}\ and\ \citenamefont
  {Desbois}(1972)}]{roulleau72}%
  \BibitemOpen
  \bibfield  {author} {\bibinfo {author} {\bibfnamefont {M.}~\bibnamefont
  {Roulleau}}\ and\ \bibinfo {author} {\bibfnamefont {M.}~\bibnamefont
  {Desbois}},\ }\href@noop {} {\bibfield  {journal} {\bibinfo  {journal}
  {Journal of the Atmospheric Sciences}\ }\textbf {\bibinfo {volume} {29}},\
  \bibinfo {pages} {565} (\bibinfo {year} {1972})}\BibitemShut {NoStop}%
\bibitem [{\citenamefont {Richardson}\ \emph {et~al.}(1989)\citenamefont
  {Richardson}, \citenamefont {Pigg},\ and\ \citenamefont
  {Hightower}}]{richardson89}%
  \BibitemOpen
  \bibfield  {author} {\bibinfo {author} {\bibfnamefont {C.}~\bibnamefont
  {Richardson}}, \bibinfo {author} {\bibfnamefont {A.}~\bibnamefont {Pigg}}, \
  and\ \bibinfo {author} {\bibfnamefont {R.}~\bibnamefont {Hightower}},\ }in\
  \href@noop {} {\emph {\bibinfo {booktitle} {Proceedings of the Royal Society
  of London A: Mathematical, Physical and Engineering Sciences}}},\ Vol.\
  \bibinfo {volume} {422}\ (\bibinfo {organization} {The Royal Society},\
  \bibinfo {year} {1989})\ pp.\ \bibinfo {pages} {319--328}\BibitemShut
  {NoStop}%
\bibitem [{\citenamefont {Taflin}\ \emph {et~al.}(1989)\citenamefont {Taflin},
  \citenamefont {Ward},\ and\ \citenamefont {Davis}}]{taflin89}%
  \BibitemOpen
  \bibfield  {author} {\bibinfo {author} {\bibfnamefont {D.~C.}\ \bibnamefont
  {Taflin}}, \bibinfo {author} {\bibfnamefont {T.~L.}\ \bibnamefont {Ward}}, \
  and\ \bibinfo {author} {\bibfnamefont {E.~J.}\ \bibnamefont {Davis}},\
  }\href@noop {} {\bibfield  {journal} {\bibinfo  {journal} {Langmuir}\
  }\textbf {\bibinfo {volume} {5}},\ \bibinfo {pages} {376} (\bibinfo {year}
  {1989})}\BibitemShut {NoStop}%
\bibitem [{\citenamefont {Betel{\'u}}\ \emph {et~al.}(2006)\citenamefont
  {Betel{\'u}}, \citenamefont {Fontelos}, \citenamefont {Kindel{\'a}n},\ and\
  \citenamefont {Vantzos}}]{betelu06}%
  \BibitemOpen
  \bibfield  {author} {\bibinfo {author} {\bibfnamefont {S.}~\bibnamefont
  {Betel{\'u}}}, \bibinfo {author} {\bibfnamefont {M.}~\bibnamefont
  {Fontelos}}, \bibinfo {author} {\bibfnamefont {U.}~\bibnamefont
  {Kindel{\'a}n}}, \ and\ \bibinfo {author} {\bibfnamefont {O.}~\bibnamefont
  {Vantzos}},\ }\href@noop {} {\bibfield  {journal} {\bibinfo  {journal}
  {Physics of Fluids}\ }\textbf {\bibinfo {volume} {18}},\ \bibinfo {pages}
  {051706} (\bibinfo {year} {2006})}\BibitemShut {NoStop}%
\bibitem [{\citenamefont {Burton}\ and\ \citenamefont
  {Taborek}(2011)}]{burton11}%
  \BibitemOpen
  \bibfield  {author} {\bibinfo {author} {\bibfnamefont {J.~C.}\ \bibnamefont
  {Burton}}\ and\ \bibinfo {author} {\bibfnamefont {P.}~\bibnamefont
  {Taborek}},\ }\href@noop {} {\bibfield  {journal} {\bibinfo  {journal}
  {Physical Review Letters}\ }\textbf {\bibinfo {volume} {106}},\ \bibinfo
  {pages} {144501} (\bibinfo {year} {2011})}\BibitemShut {NoStop}%
\bibitem [{\citenamefont {Gawande}\ \emph {et~al.}(2017)\citenamefont
  {Gawande}, \citenamefont {Mayya},\ and\ \citenamefont
  {Thaokar}}]{gawande2017}%
  \BibitemOpen
  \bibfield  {author} {\bibinfo {author} {\bibfnamefont {N.}~\bibnamefont
  {Gawande}}, \bibinfo {author} {\bibfnamefont {Y.}~\bibnamefont {Mayya}}, \
  and\ \bibinfo {author} {\bibfnamefont {R.}~\bibnamefont {Thaokar}},\
  }\href@noop {} {\bibfield  {journal} {\bibinfo  {journal} {Physical Review
  Fluids}\ }\textbf {\bibinfo {volume} {2}},\ \bibinfo {pages} {113603}
  (\bibinfo {year} {2017})}\BibitemShut {NoStop}%
\bibitem [{\citenamefont {Garzon}\ \emph {et~al.}(2014)\citenamefont {Garzon},
  \citenamefont {Gray},\ and\ \citenamefont {Sethian}}]{garzon14}%
  \BibitemOpen
  \bibfield  {author} {\bibinfo {author} {\bibfnamefont {M.}~\bibnamefont
  {Garzon}}, \bibinfo {author} {\bibfnamefont {L.}~\bibnamefont {Gray}}, \ and\
  \bibinfo {author} {\bibfnamefont {J.}~\bibnamefont {Sethian}},\ }\href@noop
  {} {\bibfield  {journal} {\bibinfo  {journal} {Physical Review E}\ }\textbf
  {\bibinfo {volume} {89}},\ \bibinfo {pages} {033011} (\bibinfo {year}
  {2014})}\BibitemShut {NoStop}%
\bibitem [{\citenamefont {Collins}\ \emph {et~al.}(2008)\citenamefont
  {Collins}, \citenamefont {Jones}, \citenamefont {Harris},\ and\ \citenamefont
  {Basaran}}]{collins08}%
  \BibitemOpen
  \bibfield  {author} {\bibinfo {author} {\bibfnamefont {R.~T.}\ \bibnamefont
  {Collins}}, \bibinfo {author} {\bibfnamefont {J.~J.}\ \bibnamefont {Jones}},
  \bibinfo {author} {\bibfnamefont {M.~T.}\ \bibnamefont {Harris}}, \ and\
  \bibinfo {author} {\bibfnamefont {O.~A.}\ \bibnamefont {Basaran}},\
  }\href@noop {} {\bibfield  {journal} {\bibinfo  {journal} {Nature}\ }\textbf
  {\bibinfo {volume} {4}},\ \bibinfo {pages} {149} (\bibinfo {year}
  {2008})}\BibitemShut {NoStop}%
\bibitem [{\citenamefont {Collins}\ \emph {et~al.}(2013)\citenamefont
  {Collins}, \citenamefont {Sambath}, \citenamefont {Harris},\ and\
  \citenamefont {Basaran}}]{collins13}%
  \BibitemOpen
  \bibfield  {author} {\bibinfo {author} {\bibfnamefont {R.~T.}\ \bibnamefont
  {Collins}}, \bibinfo {author} {\bibfnamefont {K.}~\bibnamefont {Sambath}},
  \bibinfo {author} {\bibfnamefont {M.~T.}\ \bibnamefont {Harris}}, \ and\
  \bibinfo {author} {\bibfnamefont {O.~A.}\ \bibnamefont {Basaran}},\
  }\href@noop {} {\bibfield  {journal} {\bibinfo  {journal} {Proceedings of the
  National Academy of Sciences}\ }\textbf {\bibinfo {volume} {110}},\ \bibinfo
  {pages} {4905} (\bibinfo {year} {2013})}\BibitemShut {NoStop}%
\bibitem [{\citenamefont {Deshmukh}\ and\ \citenamefont
  {Thaokar}(2012)}]{deshmukh12}%
  \BibitemOpen
  \bibfield  {author} {\bibinfo {author} {\bibfnamefont {S.~D.}\ \bibnamefont
  {Deshmukh}}\ and\ \bibinfo {author} {\bibfnamefont {R.~M.}\ \bibnamefont
  {Thaokar}},\ }\href@noop {} {\bibfield  {journal} {\bibinfo  {journal}
  {Physics of Fluids (1994-present)}\ }\textbf {\bibinfo {volume} {24}},\
  \bibinfo {pages} {032105} (\bibinfo {year} {2012})}\BibitemShut {NoStop}%
\bibitem [{\citenamefont {Lanauze}\ \emph {et~al.}(2015)\citenamefont
  {Lanauze}, \citenamefont {Walker},\ and\ \citenamefont
  {Khair}}]{lanauze2015}%
  \BibitemOpen
  \bibfield  {author} {\bibinfo {author} {\bibfnamefont {J.~A.}\ \bibnamefont
  {Lanauze}}, \bibinfo {author} {\bibfnamefont {L.~M.}\ \bibnamefont {Walker}},
  \ and\ \bibinfo {author} {\bibfnamefont {A.~S.}\ \bibnamefont {Khair}},\
  }\href@noop {} {\bibfield  {journal} {\bibinfo  {journal} {Journal of Fluid
  Mechanics}\ }\textbf {\bibinfo {volume} {774}},\ \bibinfo {pages} {245}
  (\bibinfo {year} {2015})}\BibitemShut {NoStop}%
\bibitem [{\citenamefont {Stone}(1990)}]{stone90}%
  \BibitemOpen
  \bibfield  {author} {\bibinfo {author} {\bibfnamefont {H.}~\bibnamefont
  {Stone}},\ }\href@noop {} {\bibfield  {journal} {\bibinfo  {journal} {Physics
  of Fluids A: Fluid Dynamics (1989-1993)}\ }\textbf {\bibinfo {volume} {2}},\
  \bibinfo {pages} {111} (\bibinfo {year} {1990})}\BibitemShut {NoStop}%
\bibitem [{\citenamefont {Chen}\ \emph {et~al.}(1995)\citenamefont {Chen},
  \citenamefont {Pui},\ and\ \citenamefont {Kaufman}}]{chen1995}%
  \BibitemOpen
  \bibfield  {author} {\bibinfo {author} {\bibfnamefont {D.-R.}\ \bibnamefont
  {Chen}}, \bibinfo {author} {\bibfnamefont {D.~Y.}\ \bibnamefont {Pui}}, \
  and\ \bibinfo {author} {\bibfnamefont {S.~L.}\ \bibnamefont {Kaufman}},\
  }\href@noop {} {\bibfield  {journal} {\bibinfo  {journal} {Journal of Aerosol
  Science}\ }\textbf {\bibinfo {volume} {26}},\ \bibinfo {pages} {963}
  (\bibinfo {year} {1995})}\BibitemShut {NoStop}%
\bibitem [{\citenamefont {Singh}\ \emph {et~al.}(2016)\citenamefont {Singh},
  \citenamefont {Khan}, \citenamefont {Koli}, \citenamefont {Sapra},\ and\
  \citenamefont {Mayya}}]{singh2016}%
  \BibitemOpen
  \bibfield  {author} {\bibinfo {author} {\bibfnamefont {S.}~\bibnamefont
  {Singh}}, \bibinfo {author} {\bibfnamefont {A.}~\bibnamefont {Khan}},
  \bibinfo {author} {\bibfnamefont {A.}~\bibnamefont {Koli}}, \bibinfo {author}
  {\bibfnamefont {B.}~\bibnamefont {Sapra}}, \ and\ \bibinfo {author}
  {\bibfnamefont {Y.}~\bibnamefont {Mayya}},\ }\href@noop {} {\bibfield
  {journal} {\bibinfo  {journal} {Particulate Science and Technology}\ }\textbf
  {\bibinfo {volume} {34}},\ \bibinfo {pages} {608} (\bibinfo {year}
  {2016})}\BibitemShut {NoStop}%
\end{thebibliography}%

\end{document}